\documentstyle[aps,preprint,prc,12pt]{revtex}
\newcommand{\be}{\begin{equation}}
\newcommand{\ee}{\end{equation}}
\begin{document}
\title{\bf CHAOTICITY IN VIBRATING NUCLEAR BILLIARDS}
\author {G.\,F. Burgio, M.\, Baldo, A.\, Rapisarda}
\address  {\it Istituto Nazionale
di Fisica Nucleare, Sezione di Catania}
\address  {\it   and Dipartimento
 di Fisica, Universit\'a di Catania }
\address  {\it  Corso Italia 57, I-95129 Catania, Italy }
\author {and P.\, Schuck }
\address  {\it Institut de Physique Nucl\'eaire,
Universit\'e de Grenoble}
\address  {\it 53 Avenue des Martyrs,
38026 Grenoble Cedex, France}
\date{ May 22, 1995}
\maketitle
\begin{abstract}
We study the motion of classical particles confined in a
two-dimensional "nuclear" billiard whose walls undergo
periodic shape oscillations according to a fixed multipolarity.
The presence of a coupling term in the single particle Hamiltonian
between the particle motion and the collective coordinate generates
a fully selfconsistent dynamics.
We consider in particular monopole oscillations and
demonstrate that self-consistency is essential in order to induce
chaotic single-particle motion. We also discuss the dissipative
behaviour of the wall motion and its relation with the order-to-chaos
transition in the dynamics of the microscopic degrees of freedom.
Analogous considerations can be extended to higher multipolarities.
\end{abstract}
Pacs: 24.60.Lz, 21.10.Re
% body of the paper
The question about the origin of dissipation of collective
motion in finite Fermi systems has been widely and deeply
investigated\cite{sp81}, but its solution is still an open problem.
One generally considers one- and two-body processes as the main
mechanisms which could explain this phenomenon, though their
mutual balance is currently a question of debate.
The one-body processes include i) escape of nucleons from the
collective potential well into the continuum states and ii)
collisions of nucleons with the nuclear wall generated by the common
selfconsistent mean field.
On the other hand, the two-body processes take into account collisions
between nucleons and are effective in the range of excitation
energies above the Fermi energy, because of the partial relaxation of
the Pauli principle.
In this letter we concentrate on one-body processes only and we
discuss the relationship between chaoticity at the microscopic level
and dissipation of the collective degrees of freedom.\par
In ref.\cite{bswi} Blocki {\it et al.} analyze the behavior of a gas
of several thousands non-interacting point particles enclosed in a
hard-wall container which undergoes periodic and adiabatic shape
oscillations, {\it i.e.}
the wall frequency is much smaller than a typical single particle
frequency.
Particles momenta are randomly distributed inside a Fermi sphere, and
this represents the only quantum ingredient of the model.
In coordinate space particles move on linear trajectories and collide
elastically against the walls. The
wall motion is not modified by the collisions with the particles,
therefore it keeps on oscillating with the same initial frequency
pumping energy into the gas.
For this system, the authors study the increase of the particles'
kinetic energy as a function of time.
They also compare their findings with the predictions of the
well-known "wall formula"\cite{wf}, which gives the dissipation
rate for each fixed multipolarity.
They find that for the quadrupole case the gas does not heat up,
whereas for higher multipolarities the gas kinetic energy increases
with time. They attribute the different
behavior to the fact that for low deformations the particles' motion
is regular and corresponds to an integrable situation, whereas for
higher multipolarities the scattering of segments of wall with
positive curvature
leads to divergence between trajectories and therefore
to chaotic motion.
Although their results look very interesting,
their application to the nuclear case is not straightforward because
i) the selfconsistent mean field is absent, ii) the total energy
is not conserved.\par
A step forward in this direction has been performed by Bauer
{\it et al.} in ref.\cite{bauer}. In this work the authors study
the damping of collective motion in nuclei within the Vlasov equation,
which is the semiclassical approximation to the
Time Dependent Hartree Fock (TDHF) equation. In the Vlasov equation,
which is solved by using the test-particle method\cite{wong},
selfconsistency is taken into account and the total energy is
conserved.
Moreover quantum effects like the Pauli principle are present in the
initial conditions and during the dynamical evolution
because of the Liouville's theorem.
A multipole-multipole interaction of the Bohr-Mottelson
type is adopted for quadrupole and octupole deformations.
In both cases the dynamical evolution shows a regular undamped
collective motion which coexists with a weakly chaotic single-particle
dynamics.\par
In order to clarify the relationship between chaos at the microscopic
level and damping of the collective motion, we consider a classical
version of the vibrating potential model for finite
nuclei (see {\it e.g.} ref.\cite{risc}). In this model several
non-interacting classical particles move
in a two-dimensional deep potential well and hit the oscillating
surface. Using polar coordinates, the Hamiltonian depends on a set of
$\{r_i, \theta_i \}$ variables, describing
the motion of the particles, and the collective coordinate $\alpha$.
The Hamiltonian reads
\be
\label {1}
H(r_i, \theta_i, \alpha) = \sum_{i=1}^A ({{p_{r_i}^2}\over {2m}} +
{{p_{\theta_i}^2} \over{2mr_i^2}} +
V(r_i, R(\theta_i))) + {{p_\alpha^2} \over{2M}} + {1\over 2}
M \Omega^2 \alpha^2
\ee
\noindent
being $\{p_{r_i}, p_{\theta_i}, p_{\alpha}\}$ the conjugate momenta of
$\{r_i, \theta_i, \alpha\}$.
$m = 938~MeV$ is the nucleon mass, and $M = m A R_o^2$ is the Inglis
mass, chosen proportional to the total number
of particles $A$ and to the squared billiard radius $R_o$.
$\Omega$ is the oscillation frequency of the collective variable
$\alpha$. The potential $V(r, R(\theta))$ is zero inside the billiard
and a very steeply rising function on the surface,
${V(r, R(\theta))} = {V_o \over
{(1 + exp({{R(\theta) - r} \over {a}}))}}$, with $V_o = 1500~MeV$ and
$a = 0.01~fm$.
The surface is described by
$R(\theta) = R_o (1 + \alpha P_L(cos\theta))$, where $P_L$ is
the Legendre polynomial with multipolarity $L$.
Therefore this potential couples the collective variable motion to
the particles' dynamics. The numerical simulation is based on the
solution of the Hamilton's equations
\begin{eqnarray}
\dot {r_i} & = &  {{p_{r_i}}\over {m}}~
{}~~;~~\dot{p_{r_i}}  = {{p_{\theta_i}^2}\over {m r_i^3}} -
{\partial V \over \partial r_i}  \\
\nonumber \\
\dot{\theta_i} & = & {{p_{\theta_i}}\over {m r_i^2}}~~ ;~~
\dot{p_{\theta_i}}  =  - {\partial V\over \partial R} \cdot
{\partial R \over \partial \theta_i}  \\
\nonumber \\
\dot{\alpha} & = & {p_\alpha \over M}~~~~~ ;~~
\dot{p_\alpha}  =  - M \Omega^2 \alpha - {\partial V \over
\partial R} \cdot {\partial R \over \partial \alpha}
\end{eqnarray}
We solve the Hamilton's equations with an algorithm of
Runge-Kutta type with typical time step sizes of $1~fm/c$
and the total energy is conserved with high accuracy.
In our investigation we consider the simplest case, {\it i.e.}
the monopole mode $L=0$.
Because of the spherical symmetry, $R$ does not depend on $\theta$ and
therefore each angular momentum $p_{\theta_i}$ is a constant of motion.
Another constant of motion is the total energy E. As a consequence
the problem is not integrable, because the number of constants of
motion is smaller than the number of degrees of freedom.\par
Before getting into the analysis of the monopole oscillation,
it is appropriate to study the equilibrium conditions for
our classical gas. In fact it is well known that in the realistic
nuclear case the collective motion takes place around equilibrium.
The equilibrium deformation parameter $\bar \alpha$
can be calculated by equating the mechanical pressure of the wall
$P_w = {dE \over dS} = {M \Omega^2 \over {2\pi}}
{\alpha \over {1+\alpha}}$ and the pressure
$P_p = \rho T = {A~T \over {\pi R_o^2 (1+\alpha)^2}}$
exerted by the particles,
being $\rho$ the particle density. One gets
$\bar \alpha (1 + \bar \alpha) = {2~T \over {m R_o^2 \Omega^2}}$,
the equation for the equilibrium value $\bar \alpha$.\par
We consider the wall oscillation taking place close to adiabatic
conditions.
For this purpose we impose a wall frequency smaller than the single
particle one and choose $\Omega = 0.05~c/fm$,
which corresponds to a oscillation period
$\tau_w = {2 \pi \over \Omega} \sim
125.66~fm/c$. The single particle oscillation period is equal to
$\tau_p = {2 R_o \over v}$,
being $v$ the maximum particle speed, $v = \sqrt {2T/m}$.
We choose $R_o = 6~fm$ and $T = 36~MeV$, and this
gives a single particle period $\tau_p \sim 43.3~fm/c$.\par
The equilibrium value $\bar \alpha$ corresponds of course to the
thermodinamic limit. Since we are considering a system with a finite
number of particles, the actual value of $\alpha$ will fluctuate
in time around $\bar \alpha$.
We have checked that, starting with a set of particle velocities
which follows a Maxwell distribution (approximately) and
$\alpha = \bar \alpha$,
the value of $\alpha$ varies in time, but with an average value very
close to $\bar \alpha$. To obtain a description of the system at the
statistical equilibrium, it is therefore necessary to consider the
Gibbs ensemble, {\it i.e.} a set of N copies of the system, all of
them with an initial value $\alpha = \bar \alpha$ and differing only
by the initial
microscopic distribution of particle velocities and positions
(all compatible with a maxwellian). We checked that indeed the value
$<\alpha>$ obtained by averaging among the different systems,
or "events", displays decreasing fluctuations as N increases and
stays very close to $\bar \alpha$.
After we assured that the numerical calculation produces the good
equilibrium properties\cite{bbrs},
we slightly changed the collective coordinate $\bar \alpha$
by a small amount $\delta \alpha = 0.15$, and
let both the particles and the spherical billiard evolve in time.\par
One possible way in order to investigate the role played by coupling
and see whether it can induce a chaotic dynamics, is drawing
Poincare's surface of sections either for the single particle
coordinate or for the collective variable. Unfortunately this
is impossible to perform in our case since the number of degrees
of freedom is much bigger than two. An alternative way to visualize
a chaotic behaviour is to draw the kind of plot shown in Fig.1.
There we display the final radial coordinate at a time $t$ of one
chosen particle vs. the one at $t=0$ for three cases:
a) the wall is not coupled to
the particles' motion and always oscillates at the same
frequency\cite{bswi} thus giving energy to the gas,
b) coupling is taken into account and the Hamilton's equations (2-4)
are solved respectively for b) one particle and c) ten particles.
The idea these plots are based on is the following : if the dynamics
is regular, two initially close points in space stay close even at
later times, but if the dynamics is chaotic the two points will soon
separate due to the exponential divergence induced by chaos.
In the first case this plot will show a regular curve,
whereas in the other one it displays a scattered plot.
Let us discuss the results shown in Fig.1. While the fixed spherical
billiard is a very well-known integrable case, the wall oscillation
changes the dynamics. In case a) only the angular momentum is
conserved but not the energy, therefore the system is not integrable
and we should
expect chaotic behaviour. In spite of that, we observe regularity
because the motion is adiabatic. In other words
there exist some adiabatic invariants
which guarantee regularity, though for a limited time interval.
This result confirms those of ref.\cite{bswi}.
If we omit any adiabaticity condition by increasing
the wall frequency, the dynamics
will be chaotic, as we checked explicitely. On the other hand,
the inclusion of coupling terms
substantially
modifies the dynamics even in the hypothesis of small values
of the adiabatic parameter $\tau_p / \tau_w$,
see Fig.1b) and 1c). In both cases the initially
regular curves change into scatter plots, which clearly show that
the coupling to the wall oscillation randomizes the single particle
motion. In addition, the case c)
shows the appearance of diffusive behaviour earlier than the case b),
because of the additional coupling among
the particles through the wall.\par
Scattered plots like those shown in cases b) and c) have a typical
fractal structure, and we found an uncertainty dimension $D = 0.899$
independent on time for the case b)\cite{tab,bbrs}.\par
A more quantitative way to characterize the dynamics in a chaotic
regime is
by means of the largest Lyapunov exponent $\bar \lambda$, which
gives the average divergence
rate between close trajectories in phase space\cite{tab}.
Let us denote with $d(t)$ the distance  between two phase space
trajectories, and $d_0 =d(0)$.
In Fig.2 we plot the time evolution of the ratio $d(t)/d_0$ for the
cases with one and ten particles moving inside the billiard.
By integrating two single particle trajectories which differ
at $t=0$ by $d_0 = 10^{-11}$, we calculated a phase space distance as
$d(t) = \sqrt{({\bf x_i}^{(1)} - {\bf x_i}^{(2)})^2}$,
being ${\bf x_i}  = \{r_i, \theta_i, p_{r_i}, p_{\theta_i} \}$.
The superscripts refer respectively to the trajectories (1) and (2).
This procedure is repeated several times,
in order to sample different regions of phase space with good accuracy.
For each sampling we get a curve like the one shown in Fig.2, and its
slope gives a value of the Lyapunov exponent $\lambda$.
A more accurate calculation can be done
simply by averaging $\lambda$ over several different events.
We found that $<\lambda>$ scales with the number of particles, in
particular $<\lambda_1> = 8 \cdot 10^{-3}~c/fm$ and
$< \lambda_{10} > = 8.8 \cdot 10^{-2}~c/fm$. Therefore the times for
the onset of chaoticity
$\tau = 1/<\lambda>$ are respectively equal to $\tau_1 = 125.3~fm/c$
and $\tau_{10} = 11.4~fm/c$.
We notice that relevant divergencies
on a macroscopic scale are visible at times longer than $\tau_1$ and
$\tau_{10}$, as shown in Fig.1b),1c).\par
Now let us concentrate on the dissipative properties of our system.
On the left-hand side of Fig.3 we plot the evolution of the collective
variable $\alpha$ vs. time, considering only one event, each
corresponding to the three cases already shown in Fig.1.
Since many events are needed in order to have a good global picture
of the macroscopic system (Gibbs ensemble), an average $<\alpha>$ over
4000 different
events is also considered and displayed on the right-hand side of
Fig.3.
First we analyze the behaviour of one single event. We note that
the amplitude of the collective motion is a regularly oscillating
curve
in the uncoupled case (Fig.3a), whereas an irregular pattern appears
in the coupled cases with one (Fig.3b)and ten particles (Fig.3c).
This is at variance with the results found in ref.\cite{bauer}.
Moreover in all three cases the single event does not show any
dissipation. We observe one main difference between cases b)
and c), {\it i.e.} the amplitude of $\alpha$ is bigger in b).
This happens because in our model the wall mass is proportional to the
number of particles confined inside the billiard. In case b) one
particle
is hitting the wall which has the same inertia, whereas in case
c) the wall has an inertia ten times bigger than the single particle
mass. Therefore in the latter case both the energy transfer and the
oscillation amplitude are smaller because collisions are more
elastic.\par
Let us now examine the behaviour of an ensemble of events,
each obtained by assigning random initial conditions to the particles
both in coordinate and momentum space.
While the ensemble average keeps unchanged the
uncoupled case d) - as expected -, some damping appears in
cases e) and f) corresponding respectively to the coupled motion of
one and ten particles.
In fact, if the ensemble contains only regular regions, no dissipation
is observed
in the collective motion, whereas if chaotic regions are populated
some damping shows up. In the latter case, once chaos has developed,
the particles hit the wall essentially at random, and this produces a
stochastic dephasing of the collective motion between different events.
If their number is huge, it can be shown that the dissipation rate
decays exponentially\cite{bbrs}. This behaviour is clearly shown
in the inner panel of Fig.3f), where we
display the time evolution of $<\alpha>$ up to $1600~fm/c$. We note
that the motion of the collective variable is completely damped out
for longer times and develops around equilibrium (wiggles are only
statistical errors due to the finite number of events). It is important
to remark that the damping is slower as we increase the number of
particles, and approaches zero in the limit $A \rightarrow \infty$.
We notice also that our damping times are longer than the ones
predicted by the wall formula\cite{wf}. More details are given
elsewhere\cite{bbrs}. \par
In conclusions, we have presented a novel approach based on the
solution of the Hamilton's equations for several classical particles
moving in a squared billiard,
in order to explain dissipation of the collective variable.
In our model the collective variable appears explicitely in the
Hamiltonian as an additional degree of freedom. We limited ourselves
to the study of monopole oscillations and found that the presence of
a coupling term in the single particle Hamiltonian
induces chaotic motion at microscopic level, even in adiabatic
conditions. This result is confirmed by the calculation of the
largest Lyapunov exponent.
As far as the collective coordinate is concerned, we found
irregular behaviour and no damping in the single event,
the latter point being in line with the work of Bauer {\it et al.}
\cite{bauer}. On the other hand, a whole
bunch displays dissipation because of incoherence among different
events.
This incoherence is produced by the chaotic single particle dynamics,
which makes all events belonging to the same ensemble strongly
different one from each other. Therefore we
hope to have demonstrated, although at classical level, that coupling
terms are essential in order to induce macroscopic dissipation.
Similar considerations can be extended to "nuclear" billiards with
higher multipolarities, where we expect an enhanced presence of
chaotic trajectories because of strong deformations. Further analysis
along these lines is currently in progress.
\noindent
This work was partially supported by the Human Capital and Mobility
Program of the European Community contract no. CHRX-CT92-0075.

\begin{figure}
\label{f1}
\noindent
\caption{The final radial coordinate for one particle is drawn as a
function of the initial one at different times t = 200, 400, 600, 800
fm/c. In part a) we display different plots for the case without
coupling, whereas in part b) and c) plots result from our model.
Calculations are  performed respectively with one
particle (b) and ten particles(c).}
\end{figure}
\begin{figure}
\label{f2}
\noindent
\caption{The evolution of the ratio $d/d_0$ (see text for
details) is displayed vs. time for the cases with 1 and 10 particles
moving inside the billiard (solid line). The straight line represents
a fit whose slope is the largest Lyapunov exponent.}
\end{figure}
\begin{figure}
\label{f3}
\noindent
\caption{On the left-hand side we show the behaviour of the
collective coordinate $\alpha$ as a function time for one event only.
In panel a) we display the result obtained without
coupling, whereas in panels b) and c) we show the result
from our model respectively with one and ten particles.
On the right-hand side we plot
an $\alpha$ averaged over 4000 events. The panels d), e) and
f) follow the same order as a), b) and c). The inner panel in 3f)
shows the long time behaviour.}
\end{figure}
\end{document}